\documentclass[11pt]{iopart}
\pdfoutput=1
\usepackage{amstext,amssymb,xspace}
\usepackage{graphicx}
\usepackage{xcolor}

\newcommand{\avg}[1]{\left\langle #1 \right\rangle}	
\newcommand{\dt}{{\rm d}t}

\newcommand{\eqref}[1]{(\ref{#1})}
\renewcommand{\d}[2]{\frac{{\rm d} #1}{{\rm d} #2}} 	

\usepackage{tikz}

\begin{document}

\title{Zero-range processes with saturated condensation: the steady state
and dynamics} 

\author{A.\ G.\ Thompson, J.\ Tailleur, M.\ E.\ Cates and R.\ A.\ Blythe} 

\address{SUPA, School of Physics and Astronomy, University of
Edinburgh, Kings Buildings, Mayfield Road, Edinburgh EH9 3JZ, UK}

\begin{abstract}
  {We study a class of zero-range processes in which \if{the
      number of particles that can be accommodated on a single site
      may not grow without bound. Thus}\fi the real-space condensation
    phenomenon does not occur and is replaced by a saturated
    condensation: that is, an extensive number of finite-size
    ``condensates'' in the steady state.} We determine the conditions
  under which this occurs, and investigate the dynamics of relaxation
  to the steady state. We identify two stages: a rapid initial growth
  of condensates followed by a slow process of activated evaporation
  and condensation. We analyze these nonequilibrium dynamics with a
  combination of mean-field approximations, first-passage time
  calculations and a fluctuation-dissipation type approach.
\end{abstract}

\maketitle

\section{Introduction}

Real-space condensation, whereby a finite fraction of a system's mass
accumulates within a microscopic region, is a spectacular phenomenon
that is observed in a wide range of dynamical systems. For example, it
is manifested experimentally in shaken granular gases in which
particles can diffuse between compartments~\cite
{Eggers1999,vanderMeer2004}: as the driving strength is reduced, the
sand grains cluster into a single compartment. One can also find
examples in models of macroeconomics, whereby a large fraction of the
available wealth is accumulated by a single
individual~\cite{Burda2002}, and of traffic flow, in which buses
serving a single route cluster together~\cite{O'Loan1998}.

One of the requirements for a thermodynamic condensation transition to
occur as the total density of particles is increased is the absence of
any restriction on the total number of particles that may occupy a
single site. In this work, we are interested in the case where such
unbounded growth of particle number is inhibited. This can happen
quite naturally within specific applications: for example, the
compartments in the granular gas experiments
of~\cite{Eggers1999,vanderMeer2004} are of finite size, and once they
contain more than a certain number of particles, any extra particles
may diffuse freely out of them. As we will show below, a vestige of
the condensation phenomenon may still be observed in the form of a
separation of the system into high- and low-density sites: we call
this \emph{saturated condensation}. The key questions then are: (i)
Under what conditions is saturated condensation observed?  (ii) How is
the state of saturated condensation approached dynamically from some
given initial condition?

In this work, we will provide answers to both these questions with
reference to a specific well-studied mathematical model, the
zero-range process (ZRP, see~\cite{Evans2000,Evans2005} for
reviews). This is a stochastic dynamical model in which particles
occupy sites of a lattice (or network) and hop to neighbouring sites
at a rate that depends only on the number of particles on the
departure site: it is for this reason that it is described as
`zero-range'. The model was introduced and solved for its steady-state
behaviour by Spitzer~\cite{Spitzer1970}, and it was realized a little
over ten years ago that real-space condensation~\cite{Bialas1997} is
possible at sufficiently large particle densities under certain
conditions on the hop rates~\cite{O'Loan1998}, even within a spatially
homogeneous system (see also \cite{Evans2000,Evans2005}). The dynamics
of condensation onto a single site has also been of interest.  Here,
the focus has been on the late-time coarsening of the excess mass into
a decreasing number of increasingly massive clusters, which in a
finite system ends with a process of mass exchange between the last
remaining clusters~\cite{Godreche2003,Grosskinsky2004}. In the steady
state, {for a finite system}, the condensate occasionally melts and
reforms on a new site: the timescale of this process has been the
subject of some
discussion~\cite{Grosskinsky2004,Majumdar2005,Godreche2005}.

Variants of the ZRP where mass accumulates on multiple sites---and in
particular an extensive number of sites---are comparatively little
studied. Schwarzkopf et al \cite{Schwarzkopf2008} examined the statics
and dynamics of a ZRP with transition rates chosen in such a way that
a single condensate is destabilized in favour of either a finite
number of extensive condensates, or a subextensive number of
subextensive `mesocondensates'. In both cases, one still has a finite
fraction of the mass occupying a vanishingly small fraction of the
sites in the thermodynamic limit, and thus a true condensation
transition is observed in this model.

By contrast, we consider here the case where the hop rates in the ZRP
are chosen such that the mass of condensates reaches a finite size
that does not increase with system size. This prevents a true
condensation transition, but nevertheless admits the possibility of
saturated condensation discussed above. {Since the true
  condensation limit can be approached by taking the upper limit on
  the size of a condensate to infinity, it seems clear that saturated
  condensation can have interesting consequences at a phenomenological
  level even if no formal singularity remains. This is similar to the
  equilibrium phenomenon of ``micellization'' in which attractive
  particles can form clusters whose size is limited by their packing
  geometry~\cite{safran}. An analysis of this problem as an instance
  of saturating condensation in equilibrium was offered by
  Goldstein~\cite{goldstein}.}

After recalling the definition of the ZRP in
section~\ref{sec:Presentation} and briefly reviewing the conditions
for a condensation transition in the homogeneous system
(section~\ref{sec:facto}), we present in section~\ref{sec:Saturation}
the conditions on the ZRP hop rates for saturated condensation to
occur. We then turn to the dynamics of the process. One can first ask
about the dynamics within the steady state. This involves evaporation
and formation of condensates, and in
section~\ref{sec:SteadyStateRates} we calculate the rates at which
both processes take place. The remainder of our work, presented in
section~\ref {sec:Stages}, concerns an investigation of the relaxation
to the steady state from a prescribed initial condition. Our main
finding is that this is a nontrivial, two-stage process. First, mass
rapidly accumulates on sites that are selected by the local dynamics
and in a way that depends on the initial condition. The number of such
sites typically differs from its global equilibrium value. This gives
rise to a much slower second stage in which condensates are nucleated
and evaporate as activated processes. We obtain a detailed portrait of
both stages of the relaxational dynamics through a combination of
mean-field theory, calculation of first-passage properties and an
approach reminiscent of a fluctuation-dissipation analysis. In certain
cases, these approximations agree remarkably well with stochastic
simulations. These analytical results thus constitute a more complete
account of a nonequilibrium condensation dynamics that has been
achieved so far.


\section{Presentation of the model: Steady state and condensation}
\label{sec:Presentation}

\begin{figure}[h]
  \begin{center}
    \begin{center}
      \includegraphics{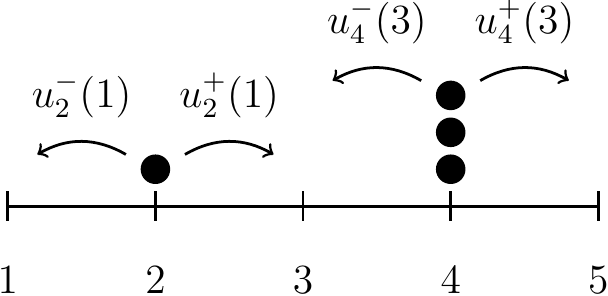}
    \end{center}\caption{Presentation of the model. The arrows indicate the
    allowed transitions and the rates at which they occur.}
  \label{fig:diagram}
  \end{center}
\end{figure}

The model we consider is defined on a one-dimensional lattice of $L$
sites with periodic boundary conditions. Each site $i$ can be occupied
by an arbitrary number of particles $n_i$. Since the system is not
connected to any reservoirs, the total number of particles in the
system $N=\sum_i n_i$ is constant. A particle can move from site $i$
to a neighbouring site $i\pm 1$ with rates $u_i^\pm(n_i)$,
respectively (see figure~\ref{fig:diagram}). We call $v_i^\pm(n)$ the
hopping rate \emph{per particle} so that $u_i^\pm(n_i)=n_i v_i^\pm
(n_i)$. Qualitatively, our {main} results hold for both symmetric and
asymmetric hoping rates but to avoid redundancies we shall only
present the symmetric case. The definition of the zero-range process
(ZRP) is that the hopping rates depend only on the number of particles
at the \emph{departure} site and not, for instance, on the occupancy
of the target site. In general the rates could also vary from site to
site but in this work we consider only spatially homogeneous systems.

\subsection{Factorization and condensation}
\label{sec:facto}

Although the ZRP has been extensively reviewed in the literature~\cite
{Evans2000,Evans2005} we shall briefly summarize some known results
that will be important in our understanding of saturated condensation,
deferring to these articles for further details. First, because the
interactions between particles are limited to a single site, the
steady-state distribution of occupancies factorizes. That is, in the
canonical ensemble one has~\cite{Evans2005}
\begin{equation}
  \label{eqn:Pss}
  P(\left\lbrace n_i \right\rbrace) = \frac 1 {Z_{N,L}}
  \prod_{i=1}^L g(n_i) \,\delta\Big(\sum_{j=1}^L n_j-N\Big)
\end{equation}
where the partition function $Z_{N,L}$ is given by
\begin{equation}
  \label{eqn:Zss}
  Z_{N,L}=\sum_{\left\lbrace
  n_i\right\rbrace}\prod_{i=1}^L g(n_i)
  \;\delta\Big(\sum_{i=1}^Ln_i-N\Big)
\end{equation}
and the factors $g(n_i)$ are determined by the hopping rates
\begin{equation}
  g(n)= \prod_{j=1}^{n} \frac 1 {u(j)} \quad \mbox{for}\quad n>0\quad \mbox{and}
\quad g(0)=1 \;.
\end{equation}
The delta functions in \eqref{eqn:Pss} and \eqref{eqn:Zss} simply
enforce the constraint that the total number of particles on the
lattice is fixed. The marginal probability that a given site has $n$
particles is given by
\begin{equation}
  p_i(n)=g(n)\frac{Z_{L-1,N-n}}{Z_{L,N}} \;.
\end{equation}
As previously mentioned, the ZRP admits an interesting condensation
transition. Although it can be worked out directly from the canonical
ensemble~\cite{EvansMZ2006}, the condition for condensation is most
easily seen in the grand canonical ensemble. Introducing a chemical
potential $\mu$ and the single-site partition function ${\mathcal
Z}_1(\mu)$, the partition function for the $L$-site system reads
\begin{equation}
  \mathcal{Z}_L(\mu)  = \sum_N\exp(\mu\,N)Z_{L,N} = \left[\sum_n {\rm e}^{\mu\, n 
-
\sum_{j=1}^{n}\ln[u(j)]}\right]^L\equiv\left[\mathcal{Z}_1(\mu)\right]^L
\end{equation}
Note that the only correlations between different sites in the
canonical ensemble come from the constraint on the total number of
particles in \eqref{eqn:Pss}. Since this constraint has been removed
in favour of a chemical potential, the sites are now completely
uncorrelated in the steady state and the $L$-site partition function
reduces to a single-site problem. To get more insight into ${\cal
Z}_1$, one can rewrite the hopping rate per particle as
$v(j)=v_0\,{\rm e}^{h(j)}$, where $v_0$ is the hopping rate of a
single particle and ${\rm e}^{h(j)}$ encodes the interaction between
the particles. For instance, $h(j)=0$ for all $j$ corresponds to
non-interacting particles. The partition function then reads
\begin{equation}
  \mathcal{Z}_1(\mu)=\sum_n \exp[-F(n,\mu)]
\end{equation}
where $F(n,\mu)=f(n)-\mu\, n$ and we have introduced
\begin{equation}
f(n)= \ln(v_0)\,n+\ln(n!)+\sum_{j=1}^{n}h(j) \;.
\end{equation}
The marginal probability that a single site has $n$ particles is then
given by
\begin{eqnarray}
  p(n|\mu) & = {\rm e}^{\mu\,n}g(n)\frac{\mathcal{Z}_{L-1}(\mu)}{\mathcal{Z}_L(\mu)} \\
  & = \frac 1 {\mathcal{Z}_1(\mu)} \exp[\mu n - f(n)]  \label{eqn:marggc}
\end{eqnarray}
and the average number of particles per site by~\cite{Evans2005}
\begin{equation}
 \left\langle n\right\rangle  \equiv \rho = \frac {\mathcal{Z}_1^\prime(\mu)}
{\mathcal{Z}_1(\mu)} \;.
\label{eq:rhoZ}
\end{equation}
We thus see from \eqref{eqn:marggc} that $f(n)$ plays the role of a
single-site free energy, {and hence that $F(n,\mu)$ is a
single-site grand canonical potential.}

The general idea of ensemble equivalence is to ask what chemical
potential $\mu$ should be imposed to get a given value of $\langle n
\rangle$. To detect a possible condensation transition, one thus looks
for the maximum density $\rho_c$ for which equation~\eqref{eq:rhoZ}
has a solution. If this maximum is infinite, then~\eqref{eq:rhoZ} is
always solvable and there is no transition.  If, however, $\rho_c$ is
finite then equation~\eqref{eq:rhoZ} cannot be solved for a density
greater than $\rho_c$ and the excess mass condenses on a single site,
which can thus carry a finite fraction of the total mass of the
system. The breaking of ensemble equivalence between canonical and
grand canonical ensemble is a signature of the condensation
transition.

The condition for \eqref{eq:rhoZ} to have a solution has been worked
out and yield a criterion on the form of $u(n)$ to observe
condensation. It can be summarized as follows~\cite{Evans2005}:
\begin{itemize}
\item if $u(n)$ decays to a non-zero constant more slowly than
$u(n)\simeq \beta (1+2/n)$, one observes above a non-zero critical
density the appearance of a single condensate in a background fluid
which remains at the critical density.
\item if $u(n)\to 0$ as $n\to \infty$, condensation occurs at all
densities and the fraction of particles in the fluid phase tends to
zero.
\item otherwise, and in particular if $u(n)$ increases as $n \to \infty$,
condensation does not occur.
\end{itemize}

The first two cases, in which there exists a true thermodynamic phase
transition, have previously received much attention in the literature
(as reviewed in \cite{Evans2005}). What we shall show in the following
is that the third case also may also exhibit interesting
condensation-like features, despite the absence of a true condensation
transition, when the stationary state supports a coexistence of sites
at two characteristic densities. We shall refer to this case as
\emph{saturated condensation}.


\subsection{Criteria for saturated condensation}
\label{sec:Saturation}

As mentioned in the Introduction, it is natural to expect condensation
in real space to saturate at some large but finite value for the mass
of the condensate (e.g., in shaken granular gases when the finite size
of the compartments prevents true condensation). We will therefore
consider systems where the hop rate per particle $v(n)$ asymptotically
decreases to a finite but non-zero value.  Even if the total hop rate
per site, $u(n)$, initially decreases with $n$, it eventually starts
increasing again and there is no phase transition, as discussed above.
However, we will show that the initial decrease in $v(n)$ may suffice
to destabilize a homogeneous state, whereas the asymptotic growth of
$u(n)$ prevents the formation of condensates with a mass that diverges
with the system size. One thus ends up with a steady-state containing
an extensive number of finite-sized condensates. Since there is no
thermodynamic transition, canonical and grand canonical ensembles
remain equivalent. Therefore, we can perform our analysis solely
within the latter ensemble, and dispense with the cumbersome
constraint on the total number of particles. On the other hand,
simulations are most straightforwardly conducted in the canonical
ensemble. Since we shall use both ensembles in the following, we will
refer equivalently to free energy or grand potential with the
understood assumption that chemical potential $\mu$ and number of
particles $N$ are adjusted so that $\sum _ n n\, p(n|\mu)=N$.

For the system to start forming condensates (i.e., high-density sites
in a sense to be defined more formally below) one needs the single
site free energy $f(n)$ to be non-convex, that is
$f^{\prime\prime}(n)<0$ for some range of $n$. Flat profiles with such
occupancies would then be unstable under the dynamics and undergo
spinodal decomposition. Treating $n$ as a continuous variable, we have
$f(n) \simeq \int_1^{n} \ln u(n') {\rm d}n'$, and hence the second
derivative with respect to $n$ is given by
\begin{equation}
  f^{\prime\prime}(n)\simeq \frac{u^{\prime}(n)}{u(n)}
\end{equation}
which implies spinodal decomposition for occupancies such that
$u^\prime(n)<0$. This is exactly equivalent to the
condition~\cite{Tailleur2008} that the hop rate per particle, $v(n) =
u(n)/n$, should satisfy the equation
\begin{equation}
  \label{eqn:critinst}
  v^{\prime}(n) < \frac {v(n)}{n} \;.
\end{equation}

Such an instability could in principle lead to complete condensation:
the criterion \eqref{eqn:critinst} is indeed satisfied when the
condensation transition occurs. This is because the phase separation
can lead to phase coexistence between a low-density phase and a
high-density phase whose density can diverge with system size, i.e., a
macroscopic condensate. This is for instance what happens
in~\cite{Schwarzkopf2008}, where the jump rates $u(n)$ depends on the
system size $L$ in such a way that the mass of each of the multiple
condensates diverges in the thermodynamic limit. For the high density
sites to have \emph{finite} occupancies, we will further require that
$f^\prime(n) \rightarrow \infty$ when ${n\rightarrow\infty}$. Combined
with the fact $f(n)$ is not everywhere convex this implies that the
free energy per site has a double tangent between two finite
densities~\cite{Rockafellar1970}. Under these conditions the grand
canonical potential per site, $F(n,\mu)$, forms a double well whose
minima occur for finite values of $n$ and give the typical occupancy
of the high- and low-density sites. `Saturated condensation' now is to
be understood as referring to this scenario. In terms of the
microscopic jump rates, the requirement that
$f^\prime(n)\rightarrow\infty$ is equivalent to
$u(n)\rightarrow\infty$ as $n\rightarrow\infty$, so $u(n)$ must be an
unbounded increasing function of $n$ or, correspondingly, $v(n)$ must
either increase as $n\to\infty$, or decrease more slowly than $1/n$.
\begin{figure}
  \centering
  \begin{minipage}{0.48\textwidth}
    \centering
    \includegraphics{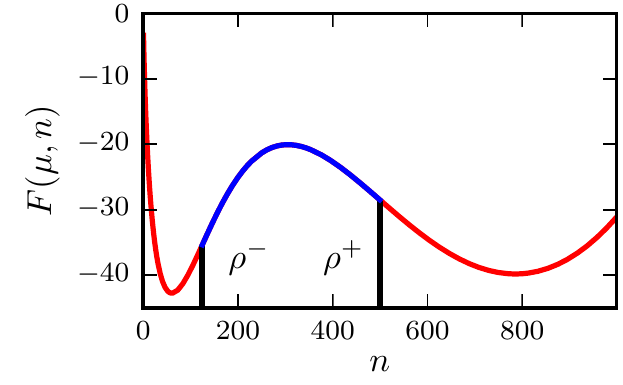}
  \end{minipage}
  \begin{minipage}{0.48\textwidth}
    \centering
    \includegraphics{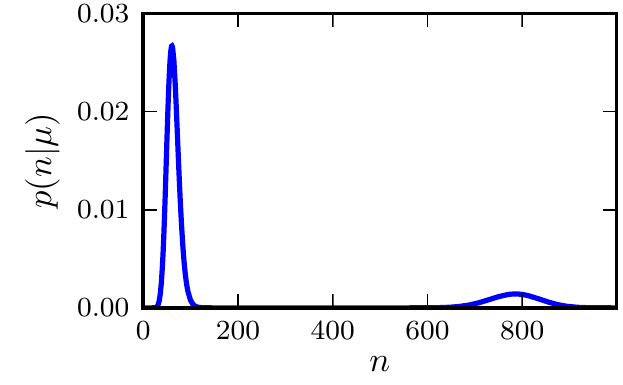}
  \end{minipage}
  \caption{{\bf Left:} The grand potential density per site,
    $F(\mu,n)$, for the choice
    $u(n)=v_0\,n\,\exp(-\lambda\phi\arctan(n/\phi))$ discussed in the
    text, with $v_0=2.5$, $\lambda=0.01$, $\phi=250$ and $\left\langle
    n\right\rangle=100$. The region unstable to spinodal decomposition
    is in blue: it corresponds to the concave part of the grand
    potential. {\bf Right:} The resulting, normalized, probability
    distribution.}
  \label{fig:ZRPFE}
\end{figure}

Let us illustrate with a concrete example of the function $u(n)$ that
leads to saturated condensation, and which we will use repeatedly
throughout the remainder of this work. It reads
\begin{equation}
  \label{eqn:ratesdefsim}
  u(n)=v_0\,n\,\exp[-\lambda\phi\arctan(n/\phi)] \;.
\end{equation}
With this definition, the hop rate per particle $v(n)$ initially
decreases exponentially with $n$ {before it saturates at a
constant value $v_0 \exp(-\lambda\phi\pi/2)$}. A flat profile is
unstable if $u'(n) <0$, that is if $\lambda \phi >2$ and
$n\in[\rho^-,\rho^+]$ where
\begin{equation}
  \rho^\pm =\frac{\lambda \phi^2 \pm \phi \sqrt{\lambda^2 \phi^2-4}}2.
\end{equation}
As $n\to \infty$, $u(n)\sim v_0 n \exp(-\lambda \phi \pi/2)$ and is
thus increasing linearly with $n$; the condensates have finite
size. This choice of rates gives rise to the double-well free energy
and bimodal probability distribution shown in fig.~\ref{fig:ZRPFE}.

\begin{figure}[h]
  \begin{center}
  \begin{minipage}{0.48\textwidth}
  \centering    \includegraphics{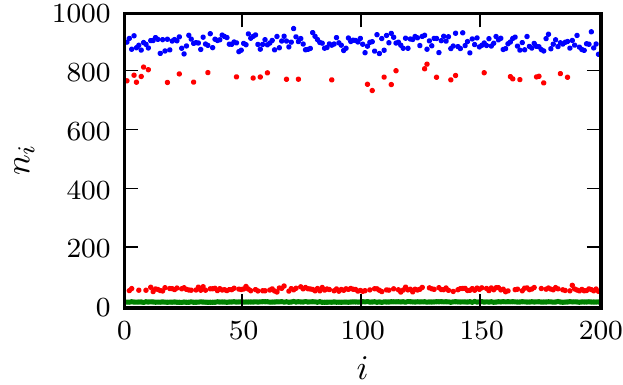}
  \end{minipage}
  \begin{minipage}{0.48\textwidth}
    \centering \includegraphics{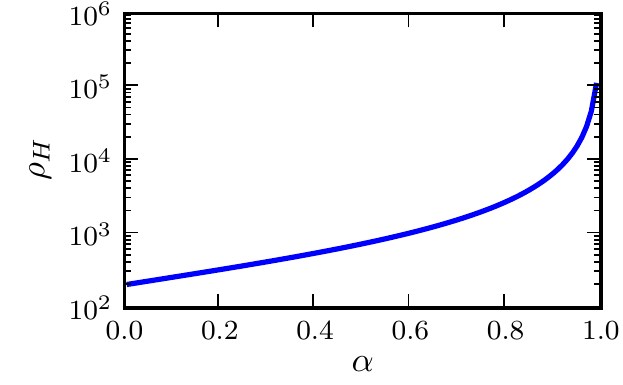}
  \end{minipage}
  \end{center}
  \caption{{\bf Left}: Steady-states of stochastic simulations for
    $\phi=250$, $\lambda=0.01$, $v_0=2.5$. For these parameters, the
    minima in the grand potential correspond to $\rho_L\approx 60$ and
    $\rho_H\approx 800$. Occupancies are averaged over a time window
    $t\in[5000;15000]$. Green and blue symbols correspond to initial
    average densities ({$\rho=20$ and $\rho=900$ respectively}) that
    are either below $\rho_L$ or above $\rho_H$ and are as expected
    stable. For an initial density {$\rho=300$} between $\rho_L$ and
    $\rho_H$ (red symbols), steady-state configurations typically
    exhibits a low density background at $\rho=\rho_L$ and high
    density condensates at $\rho=\rho_H$. {\bf Right}: Semi-log plot
    of the typical mass of the high density phase for different value
    of $\alpha$, using the rates~\eqref{eqn:valpha}. One sees that
    when $\alpha\to 1$, the mass of the condensate diverges as
    expected.}
  \label{fig:simulspinodal}
\end{figure}

Note that the range $[\rho^-,\rho^+]$ corresponds to the concave part
of the free energy, as expected from standard
thermodynamics~\cite{chaikin}. As long as the average density lies
between the two minima of $F(\mu,n)$, the steady state will be
dominated by configurations with condensates. Flat profiles will
however be metastable outside $[\rho^-,\rho^+]$, thus requiring
activated events to lead to condensation. Simulations of the systems
for rates obeying~\eqref{eqn:ratesdefsim} show the predicted
behaviour. {On the left panel of figure \ref{fig:simulspinodal} one
  sees the results of simulations started with $N$ particles
  distributed randomly over the $L$ sites. The average density
  $\rho=N/L$ is chosen either within or outside the condensation
  regime.} The criteria for the condensate to have finite mass can be
checked by considering the family of rates defined by
\begin{equation}
  \label{eqn:valpha}
  v_\alpha(n)=v_0 \, n^{-\alpha}  \exp[-\lambda\phi\arctan(n/\phi)]
\end{equation}
One indeed sees that the minimum in the grand potential corresponding
to the high-density phase is at a finite value of $n$ for $\alpha<1$
and diverges when $\alpha\to 1$ (see right panel of figure
\ref{fig:simulspinodal}).

\subsection{Condensation and evaporation dynamics in the steady state}
\label{sec:SteadyStateRates}

{When the average density lies between the two minima of the grand
  potential, there is a coexistence of high- and low-density sites in
  steady-state.} Since the grand potential barrier between them is
finite (see figure~\ref{fig:ZRPFE}), the instantaneous number of
condensates will fluctuate as low-density sites condense and
high-density sites evaporate.  To discuss these processes in more
detail, it is helpful to formally define a condensate (or high-density
site) as a site with a density greater than the one at the peak in the
grand potential between the two minima.  Likewise, when the density on
a site lies below the value corresponding to this peak, we refer it as
a low-density site. With this definition, the average densities of the
high- and low-density sites in the steady state are
\begin{equation}
  \rho_L=\frac{\sum_{n=0}^{n_{\rm peak}}
    n\,p(n|\mu)}{\sum_{n=0}^{n_{\rm peak}}p(n|\mu)}\qquad\mbox{and}\qquad
  \rho_H=\frac{\sum_{n=n_{\rm peak}}^{\infty}
    n\,p(n|\mu)}{\sum_{n=n_{\rm peak}}^{\infty}p(n|\mu)}
\end{equation}
while the average number of condensates in steady state is given by
\begin{equation}
  \label{eqn:ncss}
  n_c^{{\rm eq}}=L\,\sum_{n=n_{\rm peak}}^{\infty}p(n|\mu)= L\,\frac{\rho - 
\rho_L}
{\rho_H - \rho_L} \;.
\end{equation}

We may then also define the rate of evaporation of the condensates,
$\mathcal{R}_ {\rm evap}$, and condensation of the low-density sites,
$\mathcal{R}_{\rm cond}$.  At equilibrium these balance in such a way
that the number and size of the condensates remain constant on
average: $\mathcal{R}_{\rm evap}\,n_c^{{\rm eq}} = (L-n_c^{{\rm
eq}})\,\mathcal{R}_{\rm cond}$.

Although formation and evaporation of a condensate requires many hops
and are thus complicated processes, the corresponding rates can be
computed following a first-passage time approach~\cite{Hanggi1990}.
Indeed if we  know the first passage time
$T_{n,n_{\rm peak}}$ from a high density site $n\simeq\rho_H$ to
$n=n_{\rm peak}$, we can approximate the evaporation rate for a site
with $n$ particles by
\begin{equation}
  \mathcal{R}_{\rm evap}(n)=\frac 1 {2\,T_{n,n_{\rm peak}}} \;,
\end{equation}
where the factor of a half arises as a site at the peak can fall in
either direction with equal probability. To calculate the evaporation and 
condensation rates in the steady state, we must therefore calculate the relevant 
first-passage times for diffusion in a double well. To achieve this we follow
previous approaches applied to ZRPs undergoing a thermodynamic condensation 
transition~\cite{Godreche2005,Godreche2007}.

We illustrate this procedure by computing the first-passage time to an 
evaporation event. We thus consider a high-density site
with $n>n_{\rm peak}$ particles. The rates at which the occupancy decreases or
increases are given by
\begin{eqnarray}
  W(n\rightarrow n-1) &= u(n)=
{2\,v_0\,n}\,e^{-\lambda\,\phi\arctan\left(\frac n \phi\right)} \\
W(n\rightarrow n+1) &\equiv u_L \;.
  \label{ratesEvap}
\end{eqnarray}
In principle, the rate at which particles are added on top of a
condensate depends on the neighbouring densities. We will however
assume their fluctuations to be small and consider $u_L$ to be
constant. The first-passage time from $n$ particles to $n_{\rm peak}$ is
denoted $T_{n,n_{\rm peak}}$ and, in continuous time, is the solution to
the equation
\begin{eqnarray}
  T_{n,n_{\rm peak}} &=& \dt +
  \left[1-\left(u_L+u(n)\right)\,\dt\,\right]T_{n,n_{\rm peak}}\\ &&+
  u(n)\,\dt\, T_{n-1,n_{\rm peak}} + u_L\,\dt\, T_{n+1,n_{\rm
  peak}}\nonumber
\end{eqnarray}
Lengthy but standard algebra (see \ref{sec:solfpt}) leads to
\begin{equation}
  T_{n,n_{\rm peak}} = \sum_{l=n_{\rm peak}+1}^{n} \frac 1
{u(l)\,{p(l|\mu)}}\sum_{m=l} ^ {\infty}p({m|\mu}) \;.
  \label{FPEvap}
\end{equation}
Similarly one finds for the first passage time for condensation
\begin{equation}
  T^\prime_{n,n_{\rm peak}} = \sum_{i=n+1}^{n_{\rm peak}}\frac 1 {u_H\,p({i|\mu})} 
\sum_{j=0}^{i-1}p({j|\mu}) \;.
  \label{FPCond}
\end{equation}

\begin{figure}[t]
  \begin{center}
    \includegraphics{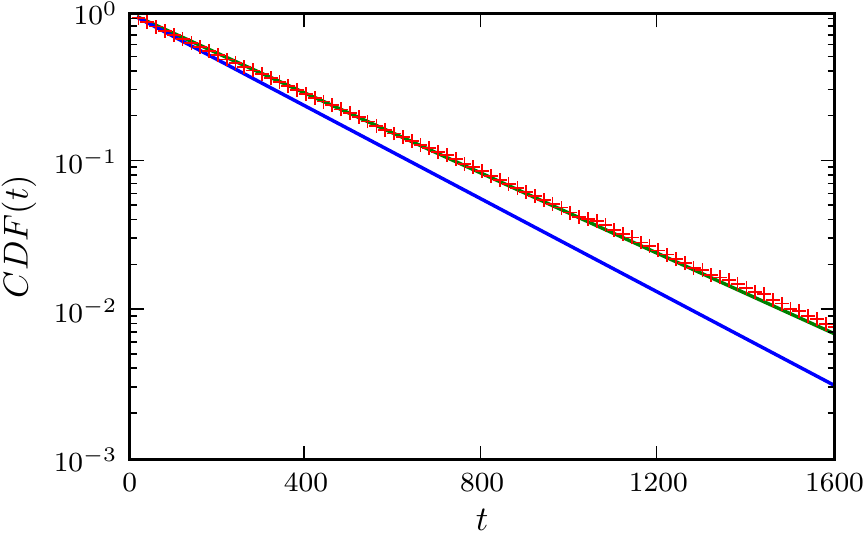}
    \caption{{Semi-log plot of the cumulative distribution function of evaporation or condensation events in steady state for $\phi=50$, $\lambda=2.8/50$, $v_0=2.5$, $\avg{n}=100$. The blue line corresponds to the cumulative distribution function of a Poisson distribution, $F_{CD}(t)=\exp(-\gamma\,t)$ with $\gamma\approx 0.0036$ as predicted by equation~\ref{eqn:TotalRates}. The red dots stem from 10000 simulations and can be fitted with a rate $\gamma\approx0.0031$ (green line).}}
    \label{fig:WaitingTime}
  \end{center}
\end{figure}

Note that to determine more accurately the rates of evaporation and
condensation we should in principle average over all starting
positions above and below the barrier, respectively:
\begin{equation}
  T_{\rm evap} = \frac{\sum_{n=n_{\rm peak}}^{\infty}
    p(n|\mu)\,T_{n,n_{\rm peak}}}{\sum_{n=n_{\rm peak}}^\infty p(n|\mu)}
  \quad\mbox{and}\quad
  T_{\rm cond} = \frac{\sum_{n=0}^{n_{\rm peak}}
    p(n|\mu)\,T^\prime_{n\,n_{\rm peak}}}{\sum_{n=0}^{n_{\rm peak}} p(n|\mu)} \;.
  \label{eqn:TotalRates}
\end{equation}
We can then define the escape rate from a configuration as the rate
for either an evaporation or condensation to occur:
\begin{equation}
  \label{eqn:R}
  \mathcal{R}_{\rm total}=\frac{n_c}{2\,T_{\rm evap}}+\frac{L-n_c}{2\,T_{\rm 
cond}} \;.
\end{equation}
In this picture, the distribution of times between events, either
evaporations or condensations, for a system of length $L$ will be
Poissonian with rate $\mathcal{R}_{\rm total}$. {For the choice of
  parameters $\phi=50$, $\lambda=2.8$, $v_0=2.5$, $L=5000$ and
  $\avg{n}=100$, equation~\eqref {eqn:R} can be evaluated numerically.} First we compute $\mu$ so that $\langle n
  \rangle = 100$ by solving \eqref{eq:rhoZ}. We can then use
  expression \eqref{eqn:marggc} for $p(n|\mu)$ to compute $T_{\rm
    evap}$ and $T_{\rm cond}$ from \eqref{FPEvap}, \eqref{FPCond} and
  \eqref{eqn:TotalRates}. We last obtain from \eqref{eqn:R} that the
  total rate is $\mathcal{R}_{\rm total}\approx0.0036$. {To compare this theoretical prediction with numerics we compute the cumulative distribution function of evaporation and condensation events from $10000$ runs: 
$F_{CD}(t)$ is the probability that the first evaporation or condensation occurs after time $t$. The simulation data are shown in
  fig.~\ref{fig:WaitingTime}; as predicted by the theory the distribution is Poissonian and a fit to
  the simulation data gives $\mathcal{R}_{\rm
    total}\approx0.0031$, which is within $10\%$ of the predicted value.}


\section{Two-stage dynamics of condensate formation}
\label{sec:Stages}

\begin{figure}[t]
  \centering
  \includegraphics{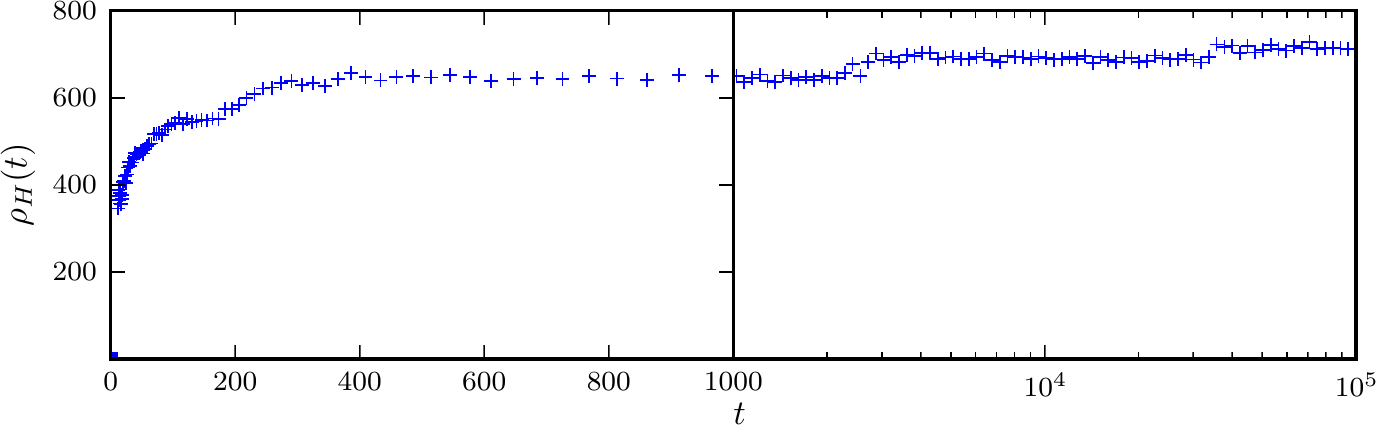}
  \caption{Average density in the high density sites as a function of
    time for parameters $\phi=250$, $\lambda=0.01$, $v_0=2.5$. The
    early dynamics ($t \sim 0-1000$) see the rapid formation of $n_c$
    condensates that are rapidly growing. The late stage dynamics
    $t\sim 10^3-10^5$ correspond to formation and evaporation of
    condensates that leads $n_c$ to $n_c^{\rm eq}$ and the average
    mass of the condensates to its equilibrium value. Steps in the
    average density correspond to the evaporation of a condensate that
    is redistributed on the surviving ones. (Note the switch from
    linear to logarithmic scale on the time axis at $t=1000$.) }
    \label{fig:illustrative} 
\end{figure}

We now examine the relaxation of the system to its steady-state, which
is a nontrivial process.  Starting from a homogeneous configuration
within the unstable region ($n\in[\rho^-,\rho^+]$, see
figure~\ref{fig:ZRPFE}), the dynamics divides naturally into two
regimes presented on figure~\ref{fig:illustrative}. The early-time
dynamics see the instability of the flat profile give birth to some
number $n_c$ of condensates that then rapidly grow. At the end of this
growth stage, $n_c$ is in general not equal to the equilibrium number
$n_c^{{\rm eq}}$ and the system has not yet reached stationarity. A
second stage then follows, taking place on much longer timescales,
during which activated events responsible for condensation and
evaporation of condensates lead the system towards its ultimate steady
state.  This difference in relevant timescales between the two stages
can be seen from figure~\ref{fig:illustrative}. Our aim in this
section is to understand these two distinct relaxational regimes.

\begin{figure}
  \centering
  \includegraphics{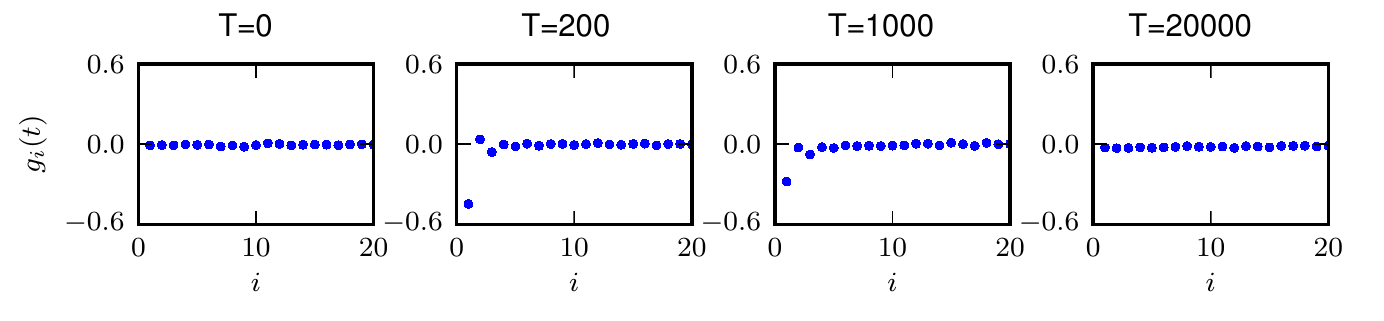}
  \caption{Snapshots of the correlation function
        $g_i(t)=\avg{n_{j+i}(t)n_j(t)}_c/\avg{n_j^2(t)}_c$, where
        $\langle x^2\rangle_c\equiv \langle x^2\rangle-\langle x
        \rangle^2$ and the averages are taken both over the lattice
        site $j$ and many simulations. Starting from an initially flat
        profile, an anti-correlation between sites forms as the
        condensates condense which then gradually disappears at late
        times when subsequent evaporations and condensations randomize
        the positions of the condensates. The parameters of the
        simulation are $v_0=2.5$, $\Phi=50$, $\lambda=.05$ and
        $\langle n\rangle=80$.}
  \label{fig:Correlation}
\end{figure}

Simulations of the stochastic system, started from random deposition
of $N$ particles over the $L$ sites of the lattice with
$N/L\in[\rho^-,\rho^+]$, show that the positions of the condensates
are, initially, anti-correlated (see figure~\ref{fig:Correlation}). 
{Transient anti-correlations of this type are a general feature of 
systems obeying a conservation law, see, for example, reference~\cite{Cornell1991}.}
This can be understood as a consequence of the condensates being
created through depletion of the neighbouring sites, thereby
preventing the formation of other condensates in their immediate
surroundings: if there is a condensate at site $i$, there is a
decrease in the probability to find another condensate in its
vicinity. These correlations survive until the late-stage dynamics
when new condensates are formed and old condensates evaporate, thus
smoothing out the correlations~\footnote{because the model is
 factorable there can be no correlations in the true steady-state}. 
Note that starting with the correct number of \emph{regularly-spaced}
condensates at the correct steady-state density would lead to
smoothing of the correlations on the same timescales: correlations are
mainly due to the immobility of the condensates, which only get
randomized (by evaporation/condensation) in the late stage of the
dynamics. Also, if one starts with a global density within
$[\rho_L,\rho_H]$ but well outside $[\rho^-,\rho^+]$, there is no
initial instability since the flat profile is metastable and the first
stage is thus absent: the creation of condensates is then only due to
activated events.  We now turn to a more detailed analysis of both
stages.

\subsection{Initial instability and growth stage}
\label{sec:PhaseI}

\begin{figure}
  \centering
  \includegraphics{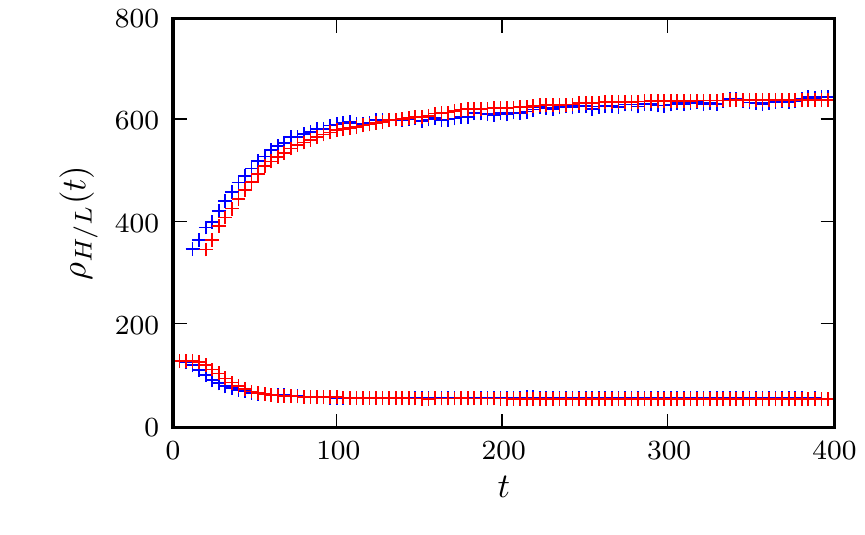}
  \caption{The evolution of the average high and low densities: using
    numerical simulations in the deterministic (red) and stochastic
    (blue) cases. Both simulations were run with random (Poissonian)
    initial conditions and with the parameters $\phi=250$,
    $\lambda=0.01$, $v_0=2.5$ and $\avg{n}=130$. Although the
    agreement is not exact the qualitative behaviour is certainly
    similar. The slight lag between the stochastic and deterministic
    cases is due to activated events increasing the initial separation
    between high and low density sites and is not especially relevant
    to an understanding of the dynamics.}
  \label{fig:MFcomparison}
\end{figure}

The early stage dynamics corresponds to the rapid growth of an
instability around the flat profile which leads to the formation of
some number of condensates $n_c$. These condensates then rapidly grow
and saturate at a density that generically differs from that at the
minimum of the high density well in the free energy. Insight into this
part of the dynamics can be gained by comparing the stochastic
dynamics of the system with its deterministic mean-field limit. The
latter is obtained by replacing $\langle u(n_i) \rangle$ by $u(\langle
n_i\rangle)=u(\rho_i)$; it reads
\begin{equation}
  \dot{\rho}_i = u(\rho_{i-1})+u(\rho_{i+1})-2\,u(\rho_i) \;.
  \label{eq:NEvDet}
\end{equation}
and can be integrated numerically using, e.g. , a simple Euler
scheme. Starting from an initial condition obtained by distributing at
random $N$ particles among the $L$ sites of the lattice, we see in
figure~\ref{fig:MFcomparison} that stochastic and mean-field dynamics
agree very well, despite the fact that the mean-field approximation
(by definition) neglects both noise and correlations. We infer from
this that activated events and spatial correlations are not very
important to understand the early-stage dynamics and we shall thus
proceed using this more analytically-tractable mean-field framework.

\begin{figure}
  \begin{center}
    \centering
    \includegraphics[viewport= 178 523 432 668,clip]{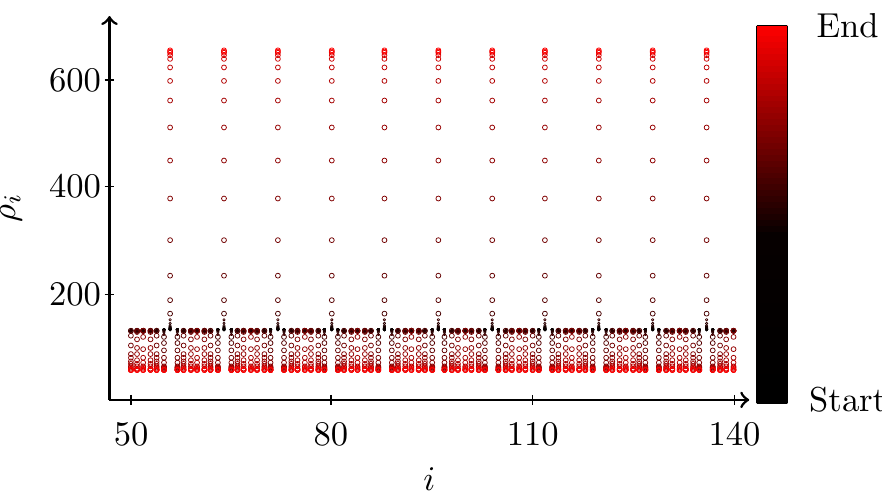}\hspace{.2cm}
    \raisebox{.3cm}{\includegraphics{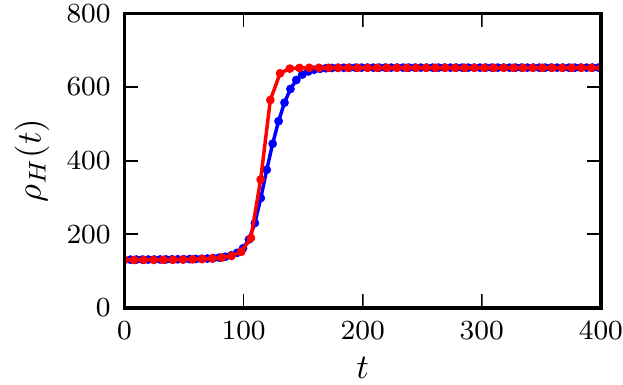}}
  \end{center}
  \caption{{\bf Left}: Starting from an initial profile which is flat
    plus a cosine perturbation condensates grow at every peak ($n_c$
    is $35\%$ higher than $n_c^{\rm eq}$). Increasing time is represented
    by a change in colour. Parameters are $\lambda=0.01$, $\phi=250$,
    $v_0=2.5$ and $\langle n \rangle =130$. {\bf Right}: The
    mean-field stochastic simulations (red), as left, and the
    solutions to equations~\eqref{eq:MFDetEv} (blue). The
    approximation to consider just two densities gives a sharper
    change but the end points are in excellent agreement.}
  \label{fig:InitialNc}
\end{figure}

At early times, the number of condensates that are created depends
strongly on the initial condition. For instance, starting the
mean-field simulation from a flat profile superposed by a cosine wave
leads to the creation of a condensate from each peak of the cosine
wave, as shown in the left panel of figure~\ref{fig:InitialNc}. We
shall first focus on the case where the number of condensates is thus
controlled. As time goes on, the mass for the condensates is drawn
from neighbouring sites, which suggests a model of this particular
condensate-formation process as one in which sites have one of two
time-dependent densities. Specifically each high-density site has a
density $\rho_H(t)$, and is surrounded by a pair of low-density sites,
both with density $\rho_L(t)$.

\begin{figure}[h]
  \centering
  \includegraphics{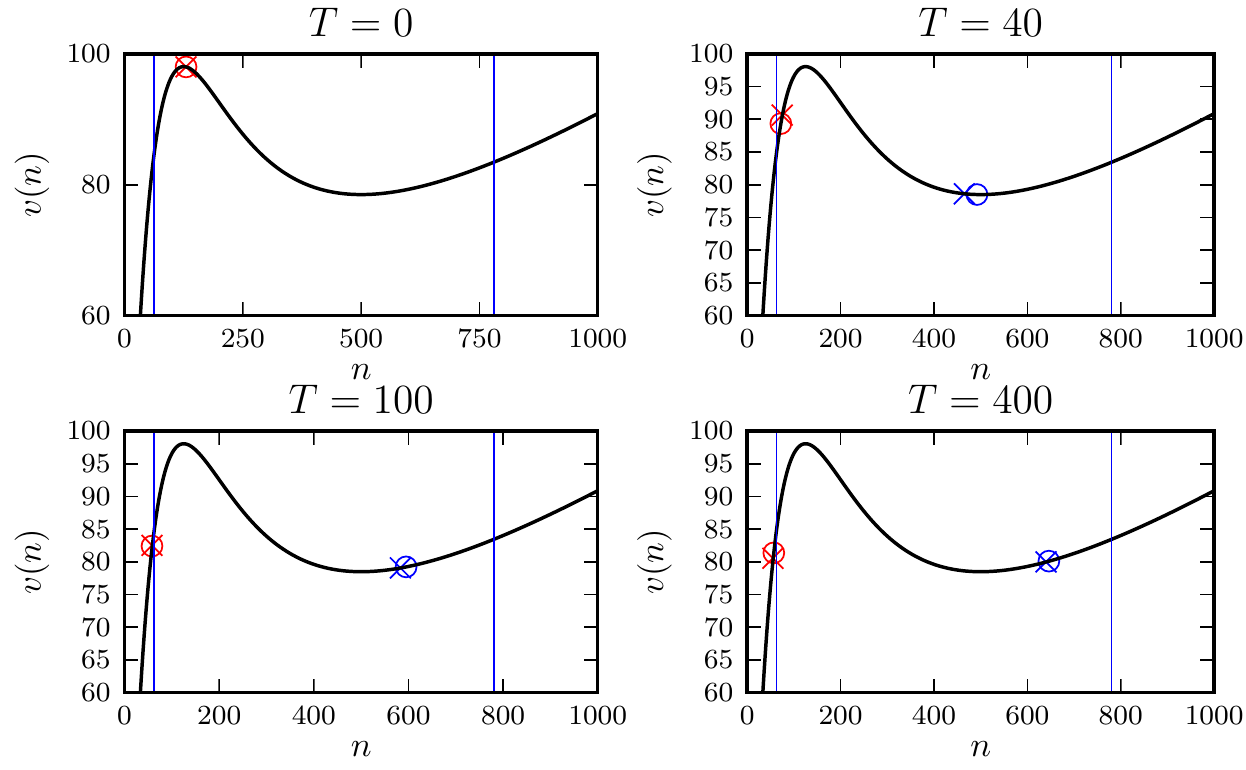}
  \caption{Snapshots of the hopping rate out of a site with $n$
    particles, $u(n)=v_0\,n\exp(-\lambda\phi\arctan(n/\phi))$, for
    mean-field simulations (crosses) and stochastic ones
    (circles). The red and blue symbols represents the average high
    density and low density sites. The blue lines show the steady
    state values of $\rho_H$ and $\rho_L$ and the black line the
    function $u(n)$. At $t=400$, $u(\rho_H)=u(\rho_L)$ and the
    condensate thus stop increasing. One must then wait for activated
    events in the stochastic simulations to get closer to the
    equilibrium values. The difference between stochastic and
    mean-field comes from the different number of condensates that
    results from the initial instability. }
  \label{fig:Vsnapshot}
\end{figure}

Within this simplification, the mean-field equations (\ref{eq:NEvDet})
reduce to
\begin{equation}
  \dot{\rho}_H = 2 \left[u(\rho_L)- u(\rho_H)\right]; \qquad
  \dot{\rho}_L = p_{\ell c}\left[u(\rho_H)- u(\rho_L)\right]
  \label{eq:MFDetEv}
\end{equation}
where we used the fact that condensates have two low density
neighbours whereas low density sites have a probability
\begin{equation}
  p_{\ell c}=\frac{2\,n_c}{L-n_c}
\end{equation}
of being next to a condensate.  This simple approximation reproduces
surprisingly well the mean-field dynamics for this particular initial
condition, as can be seen from the right panel of
figure~\ref{fig:InitialNc}. Inspection of equation \eqref{eq:MFDetEv}
then gives a simple picture of what is happening: in the spinodal
region $u^\prime(n)<0$, so that $u(\rho_L)- u(\rho_H)>0$ and
$\dot{\rho}_H$ is positive whereas $\dot{\rho}_L$ is
negative. Consequently the high density will increase and the low
density decrease. This continues as long as $u(\rho_H)<u(\rho_L)$
but stops at the first moment when $u(\rho_H)=u(\rho_L)$. This is
indeed what happens during the simulation, as can be seen on
figure~\ref{fig:Vsnapshot}.

Note that according to the previous discussion, the average values of
$\rho_L$ and $\rho_H$ at the end of the growth stage can be deduced
from the initial number of condensates, using conservation of mass and
requiring that $u(\rho_H)=u(\rho_L)$ (see equation \eqref{eq:MFDetEv})
\begin{equation}
  n_c\,\rho_H+(L-n_c)\rho_L = N \,; \qquad
  u(\rho_H) = u(\rho_L)\;.
  \label{eq:MassCons}
\end{equation}
{Starting simulations with different wavelength for the initial cosine
perturbation indeed leads to density $\rho_H$ and $\rho_L$ predicted by
\eqref{eq:MassCons} where $n_c$ equals the number of peaks of the
cosine wave.}

For more general initial conditions or in the stochastic case, we have
not been able to find a simple way to predict the number of
condensates to be formed. The saturation of their growth once formed
however follows the same rules as above, leading to high and low
densities that in general differ from the steady-state ones ({see
figure \ref{fig:Vsnapshot}}). Further changes in the average densities
are due to activated events which change the number of condensates and
increase both $\rho_L$ and $\rho_H$. These events are not captured by
the mean-field approximation, and require a distinct analysis that is
discussed in the next section.


\subsection{Activated events and late-stage dynamics}
\label{sec:Coarsening}

To understand the late stage dynamics, which is mediated by stochastic
nucleation and evaporation of condensates, we must investigate the
activated crossing between the two wells in the grand-potential
landscape shown in figure~\ref{fig:ZRPFE}.
\begin{figure}
  \centering  \includegraphics{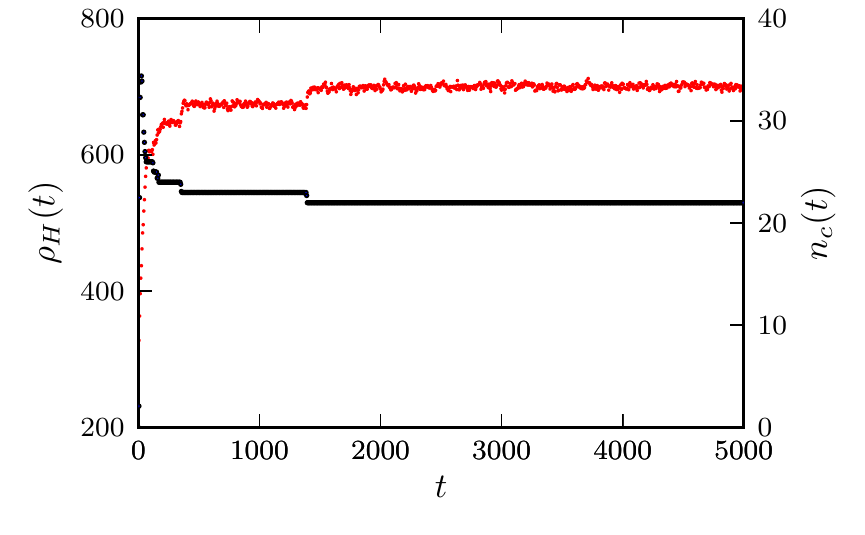}
    \caption{Typical evolution of the average density and number of
      condensate sites. Each steps correspond to the evaporation of a
      condensate that is redistributed over the other high density
      sites. The parameters of the simulations are $\lambda=0.01$,
      $\phi=250$ and $v_0=2.5$}
    \label{fig:stepwise}
\end{figure}

We have observed that during the late-stage relaxation, the total mass
in the high-density sites remains approximately constant, despite the
fact that the number of such sites changes over time. Thus, the
average density per condensate increases in the step-wise fashion
depicted in figure~\ref{fig:stepwise}.  When a condensate evaporates
the excess density on that site is redistributed over the other
high-density sites, thereby increasing their average density;
conversely, when a new condensate forms, the average density
decreases.
\if{The
time interval between steps gradually increases with each step and the
relaxation slows down as the steady state is approached, as can be
seen in fig.~\ref{fig:relaxNc}.}\fi
Depending on whether the late stage starts with too many or too few
condensates, evaporation or condensation will first dominate, before
the two rates become closer and closer. Once evaporation and
condensation of new condensates balance, the steady state described in
section \ref{sec:SteadyStateRates} is reached.

\begin{figure}
  \begin{center}
    \begin{minipage}{0.48\textwidth}
      \includegraphics{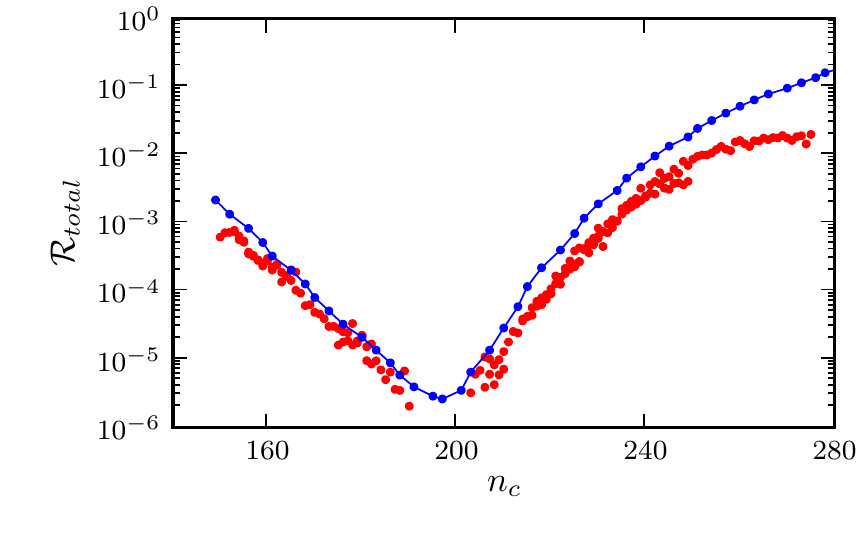}
    \end{minipage}
    \caption{Red points show data from stochastic simulations of the
      microscopic dynamics. The time spent in a configuration with $n_c$
      condensates was averaged over multiple runs and the total rate to
      leave a given configuration was taken as the inverse of this average
      time. The blue curve shows the rate calculated by using
      equations~\eqref{eqn:TotalRates} to determine the first-passage time
      to the peak and assuming a site at the peak is equally likely to fall
      into either well. All data calculated with $v_0=2.5$, $\phi=50$,
      $\lambda=3/50$, $\avg{n}=100$ and on a lattice of length $L=500$
      sites.}
    \label{fig:TotalRates}
  \end{center}
\end{figure}

In the steady state, the rates of evaporation and condensation can be
treated as a first passage problem in a grand potential landscape---as
shown previously in section \ref{sec:SteadyStateRates}. Away from
steady state, however, the rates are different from the equilibrium
ones and depend in general on the dynamics. Numerically, we can
measure the rates as a function of the fraction of condensates by
recording how long the system spends in a given configuration with
$n_c$ condensates and averaging over many runs. We now show that
these \emph{nonequilibrium} rates can also be calculated by appealing
to a fluctuation- dissipation-type argument and adapting the
equilibrium formalism correspondingly.
\begin{figure}[t]
  \begin{minipage}{0.48\textwidth}
    \centering
    \includegraphics{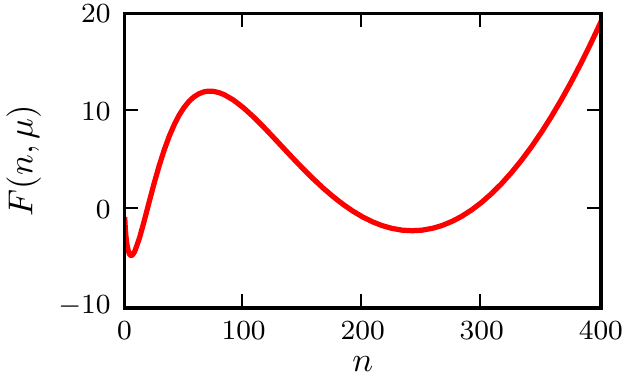}
  \end{minipage}
  \begin{minipage}{0.48\textwidth}
    \centering
    \includegraphics{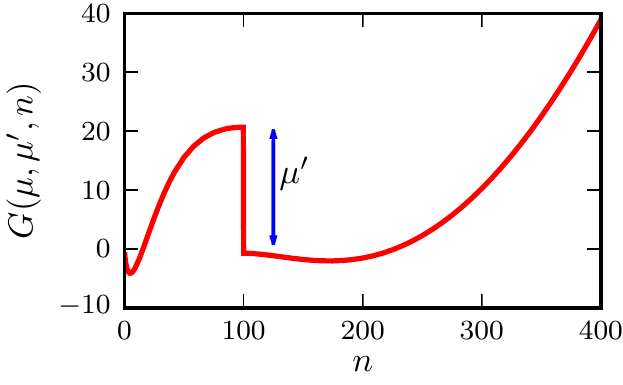}
  \end{minipage}
  \caption{{\bf Left}: the free energy per site for $v_0$=2.5,
    $\phi=50$, $\lambda=3/50$, $\avg{n}=100$ at steady state, where
    there are 40\% of condensate sites. {\bf Right}: the free energy for
    the same microscopic rates and average occupancy but constrained to
    have 60\% of condensate sites.}
  \label{fig:StepFreeEnergy}
\end{figure}
The key to this approach is to assume that when the number of
condensates $n_c$ is sufficiently close to its equilibrium value
$n_c^{\rm eq}$, the difference $n_c-n_c^{\rm eq}$ could be due either
to a spontaneous fluctuation (which we are observing) or
\emph{equivalently} to the application of a small field. To this end,
we introduce a new `doubly-grand canonical' ensemble that involves an
additional chemical potential $\mu'$ conjugate to the number of
condensates $n_c$. The corresponding partition function is then given
by
\begin{equation}
  \Xi =
\sum_{N=0}^{\infty}\sum_{n_c=0}^{L}{\rm e}^{\mu\,N}{\rm e}^{\mu^\prime\,n_c}Z_{L,N}
\;.
\end{equation}
Defining a condensate, as before, as any site containing more than
$n_{\rm peak}$ particles, the partition function can be re-written as
\begin{equation}
  \Xi  = \sum_{n_c=0}^{L}{\rm e}^{\mu^\prime\,n_c}\left[\sum_{n=0}^{\infty}{\rm e}^{-f(n)+
\mu\,n}\right]^L  = \sum_{n_c=0}^{L}\left[\sum_{n=0}^{\infty} \exp[-G(\mu,\mu',n)]\right]^L \;.
\end{equation}
where we have introduce a new thermodynamic potential 
\begin{equation}
  G(\mu,\mu',n)=f(n)-\mu\,n-\mu^\prime\,\theta(n-n_{\rm peak})
\end{equation}
$G(\mu,\mu',n)$ can simply be obtained by introducing a step $-\mu'$
in the grand potential $F(\mu,n)$, see
fig.~\ref{fig:StepFreeEnergy}. The new marginal probability to observe
an occupancy $n$ is then given by
\begin{equation}
  p(n|\mu,\mu')= \frac {{\rm e}^{-G(\mu,\mu',n)}}{ \Xi_1(\mu,\mu')} ;\quad\mbox{where}\quad \Xi_1(\mu,\mu')=\sum_{n=0}^\infty {\rm e}^{-G(\mu,\mu',n)}
\end{equation}
To evaluate the rates of evaporation and condensation of the system in
the presence of $n_c$ condensates, one can thus compute numerically
the values of $\mu$ and $\mu'$ such that
\begin{equation}
  \sum_n n \,p(n|\mu,\mu')=N\quad\mbox{and}\quad \sum_n \theta(n-n_{\rm
  peak}) \,p(n|\mu,\mu')=n_c
\end{equation}
We can then compute the rate of evaporation or condensation from a
configuration with $N$ particles and $n_c$ condensates exactly as in
section \ref{sec:SteadyStateRates} where $F(\mu,n)$ is now replaced by
$G(\mu,\mu',n)$. {To compare the results of these calculations with numerics we started 100 runs from $n_c=275$ and 100 from $n_c=155$. While these runs relaxed to the equilibrium value $n_c=199$ we recorded the average time spent, $\tau(n_c)$, by the system for each intermediate value of $n_c$ and approximated $\mathcal{R}_{\rm{total}}(n_c)=1/\tau(n_c)$. The results of these simulations are compared with
the theoretical predictions in
figure~\ref{fig:TotalRates}. The fact that the agreement between theory and numerics
is not as good as in section~\ref{sec:SteadyStateRates} may be due to 
the fact that we have poorer statistics (100 runs against 10000)}. Nevertheless, we find
that this `doubly- grand canonical' construction provides remarkably
good estimates for the evaporation and condensation rates: even where
they are two orders of magnitude larger than in equilibrium, theory
and simulations are still within a factor 2 of each other.

\section{Conclusion}
\label{sec:Conclusion}

In this work we have identified a \emph{saturated condensation}
scenario that may occur within the zero-range process (ZRP). The
steady state is characterized by an extensive number of finite-sized
condensates, as opposed to a single macroscopic condensate that has
previously been the focus of attention.  Such a state may be brought
about, for example, by a constraint on the total mass that may occupy
a single site. We have determined the conditions on the hop rates
within the ZRP that must be satisfied for saturated condensation to
arise, and have investigated various aspects of the model's dynamics,
both within and en-route to the steady state.

Our results complement existing treatments of the condensation
dynamics for the version of the ZRP with a true condensation
transition \cite{Godreche2003,
  Grosskinsky2004,Majumdar2005,Godreche2005,Schwarzkopf2008}. In those
works the focus has mostly been on the late-time scaling behaviour of
cluster size, and the characteristic timescales for evaporation and
reformation of condensates. Within the model of saturated condensation
discussed here, we have obtained a more complete account of the
dynamics---including the nontrivial behaviour of the number of
condensate sites---from very early times right through to the steady
state. In particular, we find that the relaxation takes place in two
stages: first, some number of condensate sites is dynamically selected
which depends on the initial condition. These condense rapidly, but
leave the system in an out-of-equilibrium state that slowly relaxes
through activated evaporation and condensation events.  A mean-field
approximation proved reliable in analyzing the first stage, and a
first-passage calculation conducted within a specially-constructed
grand canonical ensemble well described the second.

There are, however, some discrepancies between these approximate
analytical approaches and the full stochastic dynamics.  For example,
the rates we calculated for evaporation and condensation, both at
equilibrium and away from the steady state, seem to overestimate those
measured from the simulations. One plausible explanation for this
inconsistency is the assumption that the neighbouring sites will have
a constant density throughout the evaporation or condensation process.
In fact the neighbouring sites are likely to have higher densities
than otherwise during an evaporation and lower densities during a
condensation. As both high and low density sites sit in regions where
$v^\prime(n)>0$, an increase in density will increase the rate at
which particles enter the evaporating site whilst a decrease in
neighbouring density will decrease the rate at which particles enter a
condensing site. Both these processes will have the effect of
increasing the average duration of a condensation or evaporation
event, and reducing the rates at which they happen. Hence the
calculated evaporation and condensation rates would be larger than the
real rates as measured.

More serious, perhaps, is the absence of a satisfactory explanation
for the number of condensate sites formed from a uniform initial
condition. One would anticipate that the early-time noise would play a
major role in determining this number, although we have been unable to
relate these two quantities directly. This we leave as a possible
avenue for future work.

Finally, since we believe our model to reproduce more faithfully
experimental situations such as the shaken granular gases than the
traditional version of the ZRP, it would be very interesting to
investigate both early- and late-time dynamics of the corresponding
experiments~\cite{vanderMeer2004}.



\section*{Acknowledgements}
We thank Martin Evans and Hugo Touchette for fruitful discussions and
acknowledge funding from {\em the Carnegie Trust for the Universities
of Scotland} (A.G.T.), EPRSC EP/E030173 (J.T. \& M.E.C.) and RCUK
(R.A.B.). M.E.C. holds a Royal Society Research Professorship.


\pagebreak
\appendix
  \renewcommand{\theequation}{A-\arabic{equation}}

\section{Solution of the first passage time problem}

\label{sec:solfpt}

The first passage time from $n$ particles to
$n_{\rm peak}$ is denoted $T_{n,n_{\rm peak}}$ and, in continuous time, is
given by the solution to the equation
\begin{equation}
\label{eqn:toto}
\fl T_{n,n_{\rm peak}} = \dt +
\left[1-\left(u_L+u(n)\right)\,\dt\,\right]T_{n,n_{\rm peak}} +
u(n)\,\dt\, T_{n-1,n_{\rm peak}} + u_L\,\dt\, T_{n+1,n_{\rm peak}}
\end{equation}
This equation states that the time to go from $n$ to $n_{\rm peak}$
particles (l.h.s) is $dt$ plus the time to go from the new number of
particles, obtained after a time interval $dt$, to $n_{\rm peak}$.
With probabilities $u(n)\, dt$ and $u_L\,dt$, there are now $n-1$ or
$n+1$ particles, while with probability $1-(u_L+u(n))\,dt$ there are
still $n$ particles, hence the three terms of the r.h.s.
Equation~\eqref{eqn:toto} then reduces to
\begin{equation}
\left(u_L+u_n\right)T_{n,n_{\rm peak}} - u_L\,T_{n+1,n_{\rm peak}} - u(n)\,T_
{n-1,n_
{peak}} = 1
\end{equation}
with boundary condition $T_{n_{\rm peak},n_{\rm peak}}=0$. Now, define the
difference $d_n = T_{n,n_{\rm peak}}-T_{n-1,n_{\rm peak}}$ so that
\begin{equation}
 u(n)\,d_n-u_L\,d_{n+1} = 1.
\label{diffEvap}
\end{equation}
Note that the evolution of the probability to find $n$ particles at a
site, $p(n|\mu) $, is given by
\begin{eqnarray}
  \d{p(n|\mu)}{t} & = u(n+1)\,p(n+1|\mu)+u_L\,p(n-1|\mu) -(u(n)+u_L)\,p(n|\mu)\\
  & \equiv J_{n+1,n} - J_{n,n-1}
\end{eqnarray}
where $J_{n+1,n}=u(n+1)\,p(n+1|\mu) - u_L\,p(n|\mu)$. At steady state the left
hand side of this equation must equal zero, so the current, $J$ must
be constant. However, as $p(n<0|\mu)=0$ and $v_0=0$, $J_{-1,0}=0$ and
hence this constant must be zero. This implies that $u(n+1)\,p(n+1|\mu) =
u_L\,p(n|\mu)$ so the solution to the homogeneous version of
equation~\eqref{diffEvap} is:
\begin{equation}
d_n = \frac 1 {u(n)\,p(n|\mu)}.
\end{equation}
To solve the inhomogeneous equation, then, we look for solutions of the form 
\begin{equation}
d_n = \frac {c_n} {u(n)\,p(n|\mu)}.
\end{equation}
where $c_n$ is to be determined. Substituting this expression back
into equation~ \eqref{diffEvap} gives
\begin{equation}
-u_L\,\frac{c_{n+1}}{u(n+1)\,p(n+1|\mu)}+u(n)\,\frac{c_n}{u(n)\,p(n|\mu)} = 1.
\end{equation}
The detailed balance condition, $u(n+1)\,p(n+1|\mu) = u_L\,p(n|\mu)$,
then implies
\begin{equation}
c_n -c_{n+1} = p(n|\mu),
\end{equation}
so that $c_n = \sum_{l=n}^{\infty}p(l|\mu)$ and the first passage time is given by
\begin{equation}
T_{n,n_{\rm peak}} = \sum_{l=n_{\rm peak}+1}^{n} \frac 1 {u(l)\,p(l|\mu)}\sum_{m=l}^
{\infty}p(m|\mu).
\label{FPEvap2}
\end{equation}

\end{document}